\documentclass[fleqn,usenatbib]{mnras}
\usepackage{xcolor}
\usepackage{amsmath,amssymb,bm}
\usepackage{graphicx}

\title[Laplace surface around Kerr black hole]{Laplace surface of eccentric orbits around Kerr black hole and black-hole effects on Lidov-Kozai cycles}

\author[]{Haonan Quan\thanks{Email: 2849453104@qq.com}, Xing Wei\thanks{correspondence author: xingwei@bnu.edu.cn}\\
School of Physics and Astronomy, Beijing Normal University, Beijing, 100875, P.R.China}
\date{Accepted XXX. Received YYY; in original form ZZZ}
\pubyear{\the\year{}} 

\begin{document}

\label{firstpage}
\pagerange{\pageref{firstpage}--\pageref{lastpage}}

\maketitle

\begin{abstract}
We study the Laplace surface of eccentric orbits around a Kerr black hole with a companion star. The two relativistic effects (periapsis apsidal precession and  Lense-Thirring  precession) balance companion's tidal effect. By solving the secular equations, we find that the two branches of equilibrium solutions exist only in a narrow range of orbital inclinations of test particle, and that the stability exists when companion's obliquity approaches 90 degree. We then study the black-hole effects on Lidov-Kozai cycles. In addition to the black-hole suppression on cycle amplitude, we find that the cycle frequency is non-monotonic about distance from black hole and reaches its minimum in the transition between GR apsidal precession and companions's tidal precession. Finally we use this mechanism to well interpret the nearly isotropic orientation of S-cluster in the Galactic center.
\end{abstract}

\begin{keywords}
celestial mechanics -- quasars: supermassive black holes
\end{keywords}

\section{Introduction}

In planetary systems, the long-term evolution of satellite orbits around an oblate planet is governed by the balance between the planet’s equatorial bulge (quadrupole moment) and the tidal field of a distant star \citep{laplace1829_bowditch}. Under these competing torques, there exists a special set of circular orbits, called the {\it Laplace surface}, on which the net secular torque vanishes, so that orbits maintain fixed orientation and shape. The classical Laplace surface transitions from the planet’s equatorial plane at small radii to its orbital plane at large radii. An extension of classical Laplace surface has been explored by Tremaine explicitly \citep{tremaine2009satellite} in which the eccentric orbit and its instability were systematically studied.

\begin{figure}
\centering
\includegraphics[width=0.46\textwidth]{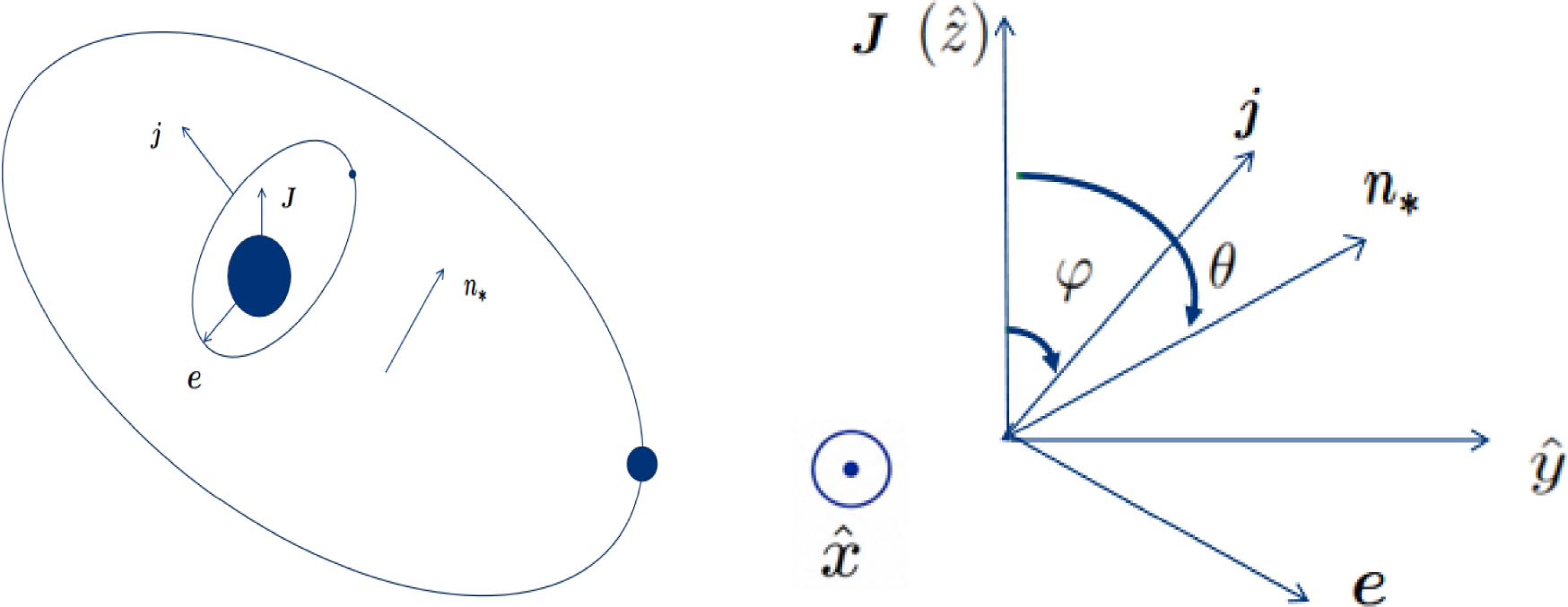}
\caption{Left: system configuration. Right: adopted frame: $\boldsymbol{J}\parallel\hat{z}$, $\boldsymbol{n}_*$ in $y-z$ plane. At equilibrium, $\boldsymbol{j}$ and $\boldsymbol{e}$ lie in the same plane. $\theta$: spin--star misalignment; $\varphi$: spin--orbit misalignment.}\label{fig:sketch}
\end{figure}

Here we investigate a relativistic analogy shown in Figure~\ref{fig:sketch}, a test particle orbiting a Kerr black hole with a companion star. In this case, the black hole’s spin (frame-dragging) is analogous to the planetary quadrupole.  Tilted orbits around the spinning black hole undergo Lense–Thirring  precession, which competes against the stellar tidal torque \citep{tremaine2014dynamics}. Previous studies of warped accretion disks (e.g., the Bardeen–Petterson effect) have shown that such torques can cause disk alignment and warping in X-ray binaries and AGN \citep{bardeen1975lense}, and furthermore the possible disk breaking \citep{Nealon2022}. However, the previous studies focus on circular orbits around black hole but eccentric orbits are not well studied, and we will find both equilibrium and stability of eccentric orbits.

In this paper, we derive the orbit-averaged (secular) equations including both Newtonian tidal torques and relativistic corrections.  We then solve for the equilibrium conditions of the orbital angular momentum $\boldsymbol{j}$ and eccentricity $\boldsymbol{e}$ (the generalized Laplace equilibria).  We find that equilibria exist only in a narrow inclination range and belong to two solution branches.  A new characteristic radius $r_M$ naturally arises in addition to the classical Laplace radius $r_L$.  We also examine the linear stability about equilibria. These results generalize the classical Laplace surface around Kerr black hole and may have applications to warped disks and stellar orbits in relativistic case.

The paper is organized as follows.  In Section \ref{sec:equations} we derive the secular equations of motion, including the effects of the companion’s tidal field and the black hole’s spin.  In Section \ref{sec:results} we obtain the equilibrium of Laplace surface and analyze its linear stability. In Section \ref{sec:application} we apply this mechanism to orbital isotropy of S-cluster. In Section \ref{sec:summary} we give a brief summary and some discussions.

\section{Secular equations}\label{sec:equations}

We study the secular motion of a test particle orbiting a Kerr black hole with a stellar companion. For a test particle around a point mass $M$ with orbital semi-major axis $a$ and eccentricity $e$, the angular momentum $\boldsymbol{L} = \boldsymbol{r}\times \boldsymbol{v}$ and the eccentricity vector $\bm e$ conserve. We introduce the dimensionless angular momentum $\bm j=\bm L/\Lambda=\sqrt{1-e^2}\,\hat{\boldsymbol{L}}$ where $\Lambda=\sqrt{GMa}$ is the circular angular momentum. When a distant companion star of mass $M_*$ orbits with semi-major axis $a_*$ and eccentricity $e_*$, it induces a tidal potential on the particle, which leads to the quadrupole Hamiltonian $H_*(\bm r,\bm p)$ where $\bm r$ and $\bm p$ are particle's position vector and momentum. Averaging over both the orbits of test particle and companion star yields the secular Hamiltonian \citep{tremaine2023dynamics}
\begin{equation}
    \langle H_* \rangle = \frac{G M_* a^2}{8\,a_*^3(1-e_*^2)^{3/2}}
    \Big[15(\boldsymbol{e}\cdot\hat{\boldsymbol{n}}_*)^2 - 6e^2 + 1 - 3(\boldsymbol{j}\cdot\hat{\boldsymbol{n}}_*)^2 \Big],
\end{equation}
where $\hat{\boldsymbol{n}}_*$ is the unit vector normal to the companion's orbital plane. 

We describe the motion of a test particle with an osculating Keplerian orbit: at each instant the test particle follows a Keplerian orbit with the same position and velocity.  Due to the perturbations, the orbital elements of the osculating orbit, e.g., $\boldsymbol{j}$ and $\boldsymbol{e}$, slowly evolve over time, while the semi-major axis $a$ remains constant because of energy conservation. Therefore, the time evolution of $\boldsymbol{j}$ and $\boldsymbol{e}$ determines the secular motion of the test particle. Substituting $\langle H_*\rangle$ into Milankovitch equations (see Appendix~\ref{appendix}) we write the secular precession due to the companion \citep{tremaine2023dynamics}
\begin{equation}\label{eq:star-secular}
\begin{aligned}
\frac{d\boldsymbol{j}}{dt} &= \frac{3}{4}\sqrt{\frac{G}{M}}\frac{M_*a^{3/2}}{a_*^3(1-e_*^2)^{3/2}}
\Big[ (\boldsymbol{j}\cdot\hat{\boldsymbol{n}}_*)\,\boldsymbol{j}\times \hat{\boldsymbol{n}}_*
 \\
 &-5(\boldsymbol{e}\cdot\hat{\boldsymbol{n}}_*)\,\boldsymbol{e}\times \hat{\boldsymbol{n}}_*\Big],\\
\frac{d\boldsymbol{e}}{dt} &= \frac{3}{4}\sqrt{\frac{G}{M}}\frac{M_*a^{3/2}}{a_*^3(1-e_*^2)^{3/2}}
\Big[ (\boldsymbol{j}\cdot\hat{\boldsymbol{n}}_*)\,\boldsymbol{e}\times \hat{\boldsymbol{n}}_*
\\
&-5(\boldsymbol{e}\cdot\hat{\boldsymbol{n}}_*)\,\boldsymbol{j}\times \hat{\boldsymbol{n}}_* 
 +2\,\boldsymbol{j}\times \boldsymbol{e}\Big].
\end{aligned}
\end{equation}

On the other hand, in the weak-field limit, the action of a test particle around a Kerr black hole with mass $M$ and angular momentum $\bm J$ can be expanded with the post-Newtonian corrections as
\begin{equation}\label{eq:black-hole-action}
\begin{split}
S &= -mc \int_{0}^{s_{0}} ds
   = -mc^{2} \int_{0}^{t_{0}} 
   \Biggl[\, 1 - \left( \frac{v}{c} \right)^{2} 
   + \frac{2GM}{c^{2}r}  \\
   &\quad + \left( \frac{v}{c} \right)^{2} \frac{2GM}{c^{2}r} 
   + \chi \left(\frac{v}{c}\right)^{5}
   + O\!\left( \frac{v}{c}\right)^{6} \,\Biggr]^{1/2} dt
\end{split}
\end{equation}
where $\chi=Jc/(GM^2)$ is the dimensionless spin parameter. We can see the two general relativistic corrections:  $O(v^4/c^4)$ (Schwarzschild perihelion precession) and $O(v^5/c^5)$ (Lense–Thirring precession). Following Landau's method \citep{landau1975classical} we write the perihelion precession (1 PN) equation
\begin{equation}\label{eq:black-hole-M}
 \frac{d\boldsymbol{e}}{dt} = \frac{3 G M\mathbf{}{n}}{c^2 a (1 - e^2)^{3/2}} \boldsymbol{j} \times \mathbf{e},  \quad 
\end{equation}
where n is the mean frequency and the Lense–Thirring precession (1.5 PN) equations
\begin{equation}\label{eq:black-hole-J}
\begin{aligned}
\frac{d \boldsymbol{j}}{dt} &= \frac{2GJ}{c^2a^3(1-e^2)^{3/2}}\,\hat{\boldsymbol{J}}\times \boldsymbol{j},\\
\frac{d \boldsymbol{e}}{dt} &= \frac{2GJ}{c^2a^3(1-e^2)^{3/2}}\Bigl[\hat{\boldsymbol{J}} - (\hat{\boldsymbol{J}}\cdot \boldsymbol{j})\frac{\boldsymbol{j}}{1-e^2}\Bigr]\times \boldsymbol{e}.
\end{aligned}
\end{equation}

We now write the secular equations with the both effects arising from companion star and black hole. Combining Eqs.~\eqref{eq:star-secular}, \eqref{eq:black-hole-M} and \eqref{eq:black-hole-J} and following Tremaine's method \citep{tremaine2009satellite}, we are led to the dimensionless equations about $\bm j$ and $\bm e$
\begin{equation}
\label{eq:j-precession}
\begin{aligned}
 \frac{d\boldsymbol{j}}{d\tau}& = \frac{3\epsilon_*}{4}(\boldsymbol{j}\cdot\hat{\boldsymbol{n}}_*)\,\boldsymbol{j}\times \hat{\boldsymbol{n}}_*
 - \frac{15\epsilon_*}{4}(\boldsymbol{e}\cdot\hat{\boldsymbol{n}}_*)\,\boldsymbol{e}\times \hat{\boldsymbol{n}}_* \\
 &+ 2\epsilon_J\,\hat{\boldsymbol{J}}\times \boldsymbol{j},
 \end{aligned}
 \end{equation}
\begin{equation}
\label{eq:e-precession}
\begin{aligned}
\frac{d\boldsymbol{e}}{d\tau}&= \, \frac{3\epsilon_*}{4}(\boldsymbol{j}\cdot\hat{\boldsymbol{n}}_*)\,\boldsymbol{e}\times \hat{\boldsymbol{n}}_*
- \frac{15\epsilon_*}{4}(\boldsymbol{e}\cdot\hat{\boldsymbol{n}}_*)\,\boldsymbol{j}\times \hat{\boldsymbol{n}}_* \\
& + \left(\frac{3}{2}\epsilon_* + 3\epsilon_M\right)\boldsymbol{j}\times \boldsymbol{e}
+ 2\epsilon_J\,\hat{\boldsymbol{J}}\times \boldsymbol{e}\\
&- \frac{6\epsilon_J}{1-e^2}(\boldsymbol{j}\cdot\hat{\boldsymbol{J}})\,\boldsymbol{j}\times \boldsymbol{e}. 
\end{aligned}
\end{equation}
Here the dimensionless parameters are
\begin{equation}
\begin{split}
\epsilon_* &= \frac{M_* a^3}{M a_*^3 (1-e_*^2)^{3/2}}, \quad
\epsilon_J = \frac{G^{1/2}J}{M^{1/2}c^2 a^{3/2}(1-e^2)^{3/2}}, \\
\epsilon_M &= \frac{GM}{a c^2(1-e^2)^{3/2}}, \quad
\tau = \sqrt{\frac{GM}{a^3}}\,t .
\end{split}
\end{equation}
In the above equations the relations $\boldsymbol{j}\cdot\boldsymbol{e}=0$ and $\boldsymbol{j}^2+\boldsymbol{e}^2=1$ always hold. 

To search the Laplace surface in equilibrium, we require the time derivatives $d\bm j/d\tau$ and $d\bm e/d\tau$ on LHS to vanish, i.e., all the secular effects on RHS cancel each other. To evaluate the orbital stability, we perturb the equilibrium $(\boldsymbol{j},\boldsymbol{e})=(\boldsymbol{j}_0,\boldsymbol{e}_0)$ by $\boldsymbol{j}=\boldsymbol{j}_0+\boldsymbol{j}_1$, $\boldsymbol{e}=\boldsymbol{e}_0+\boldsymbol{e}_1$ with $|\boldsymbol{j}_1|,|\boldsymbol{e}_1|\ll1$. Linearizing \eqref{eq:j-precession} and \eqref{eq:e-precession} yields a set of first-order perturbation equations for $(\boldsymbol{j}_1,\boldsymbol{e}_1)$ (see Appendix~\ref{appendix}). Assuming perturbations $\propto e^{\lambda \tau}$, we can obtain an eigenvalue problem. The equilibrium is linearly stable if all eigenvalues $\lambda$ have non-positive real parts.

\citet{tremaine2014dynamics} once studied the equilibrium configuration of a warped fluid disk around a black hole, analogous to the classical Laplace surface but restricted to circular orbits ($e=0$). The Laplace surface of eccentric orbits in our model does not reduce to \citep{tremaine2014dynamics} in the limit $e\to 0$. The right-hand-side of the evolution equation for $\boldsymbol{e}$ (Eq.~\ref{eq:e-precession}) is proportional to $e$. Although setting $e=0$ satisfies the equation automatically, the equilibrium of eccentric orbits imposes constraint that will be absent in the circular orbits. If the test particles were replaced by a viscous fluid then the equilibrium configuration would recover the warped-disk solution of \citep{tremaine2014dynamics}.

\section{Equilibrium and stability of Laplace surface}\label{sec:results}

For simplicity we choose the Cartesian coordinate system with the black-hole spin $\hat{\boldsymbol{J}}$ along the $z$-axis. Without loss of generality, we restrict the companion’s orbital angular momentum $\bm n_*$ to lie in the $y$-$z$ plane defined as the principle plane (see Figure~\ref{fig:sketch}), denoting by $\theta$ the angle between $\bm n_*$ and $\hat{\bm J}$. We denote by $\varphi$ the angle between $\boldsymbol{j}$ and $\hat{\boldsymbol{J}}$.

Taking the dot product $\hat{\mathbf{n}}_\ast\cdot$ on RHS of Eq.~(\ref{eq:j-precession}) yields \((\hat{\mathbf{n}}_\ast\times \hat{\mathbf{J}})\cdot \mathbf{j}=0\), implying that $\mathbf{j}$ lies in the principal plane. Similarly, applying the same operation to Eq.~(\ref{eq:e-precession}) shows that $\mathbf{e}$ must also lie in the principal plane. Therefore, $\bm j$ and $\bm e$ are both in the principal plane and perpendicular to each other. The vectors are then described as $\boldsymbol{j}=(0,\sin\varphi,\cos\varphi)$, $\boldsymbol{e}=(0,\cos\varphi,-\sin\varphi)$, $\boldsymbol{J}=(0,0,1)$ and $\hat{\boldsymbol{n}}_*=(0,\sin\theta,\cos\theta)$. Equations (\ref{eq:j-precession})--(\ref{eq:e-precession}) then reduce to
\begin{equation}
\left(\frac{a}{r_L}\right)^{9/2}
= \frac{1}{(1+4e^2)(1-e^2)}
\frac{\sin\varphi}{\sin 2(\varphi-\theta)},
\label{eq:eq-cond1}
\end{equation}
\begin{equation}
\begin{split}
&\left(\frac{a}{r_L}\right)^{9/2}(1-e^2)(\sin^2(\varphi-\theta)-\frac{1}{4})\\
&-(\frac{a}{r_{L}})^{1/2}(1-e^2)^{1/2}\left(\frac{r_M}{r_L}\right)^{4}
+\frac{1}{4}\cos{\varphi}=0.
\label{eq:eq-cond2}
\end{split}
\end{equation}
Here, in addition to the classical Laplace radius, a new typical radius $r_M$ emerges. The two radii are defined as 
\begin{equation}
\begin{split}
    r_{L} &= \left( \frac{16}{3} \right)^{2/9}
            \left( \frac{1}{2} \right)^{1/3}
            \chi^{2/9}
            \left( \frac{M_{B}}{M_{\ast}} \right)^{2/9}
            \left( r_{S} \tilde{a}^2_{\ast} \right)^{1/3}, \\
    r_{M} &= \left( \frac{1}{2} \right)^{1/4}
            \left( \frac{M_{B}}{M_{\ast}} \right)^{1/4}
            \left( r_{S} \tilde{a}^3_{\ast} \right)^{1/4}    
\end{split}
\end{equation}
where $r_S$ is the Schwarzschild radius of the black hole, $\tilde{a}_\ast = a_\ast(1-e_\ast^2)^{1/2}$ is the effective semi-major axis of the companion. We can find that 
\begin{equation}
\frac{r_M}{r_L} \sim 
\left(\frac{M_B}{M_\ast}\right)^{1/36}
\chi^{-2/9}\left(\frac{\tilde{a}_\ast}{r_S}\right)^{1/12} \gtrsim 1.
\end{equation}
With the representative parameters, we estimate the ratio
$r_M/r_L$. For dimensionless spin \(\chi\in[0.1,0.9]\),
 black-hole mass \(M_{B}\in[10^{6}M_{\odot},\,10^{9}M_{\odot}]\), companion mass \(M_{\ast}\in[1^{}M_{\odot},\,10^{2}M_{\odot}]\), and companion semi-major axis
\(\tilde a_{\ast}\in[10\ \mathrm{au},\,2\times10^{5}\ \mathrm{au}]\), the ratio
\(r_M/r_L\) falls in the interval \([1.17,\,7.26]\).

The Lense-Thirring precession is essential for the existence of Laplace surface in our setup. In the absence of black-hole spin ($\chi=0$), the black hole drives the evolution of $\boldsymbol{e}$ and the companion drives the precession of $\bm e$ and $\boldsymbol{j}$, so that the equilibrium about $\bm j$ cannot be achieved and the system cannot admit a stationary Laplace surface. In this case, the relativistic effect can suppress the Lidov-Kozai cycles \citep{liu2015suppression}. Moreover, our setup is different from the hierarchical triple system with the Lidov-Kozai cycles driven by a distant companion \citep{munoz2015survival,grishin2018chaotic}. In a triple system the Lidov-Kozai cycles induced by a distant star on the inner binary competes against the tidal or relativistic effect. In our setup the test particle is influenced by the black hole and the companion both of which are fixed.

It is instructive to compare the equilibrium in our setup with the classical Laplace surface. In the classical problem, the planet is the central body and the perturbing star is considered as the outer companion. In our setup, the black hole takes the central role and the companion star provides the tidal perturbation. The important difference results from the perturbing potentials. For a planet, the quadrupole moment drives precession of both $\boldsymbol{e}$ and $\boldsymbol{j}$ and the octupole moment drives chaotic Lidov-Kozai cycles \citep{katz2011}. For a black hole, the relativistic effects work on quite different scales, i.e., the Schwarzschild precession at $\mathcal{O}(v^4/c^4)$ and the Lense-Thirring precession at $\mathcal{O}(v^5/c^5)$, but neither can induce Lidov-Kozai cycles. The tidal potential of companion star must simultaneously balance the two relativistic effects at two different scales, and consequently, the equilibrium will be restricted to a narrow range (Eqs.~\eqref{eq:eq-cond1}--\eqref{eq:eq-cond2}).

\begin{figure*}
\centering
\includegraphics[width=0.45\textwidth]{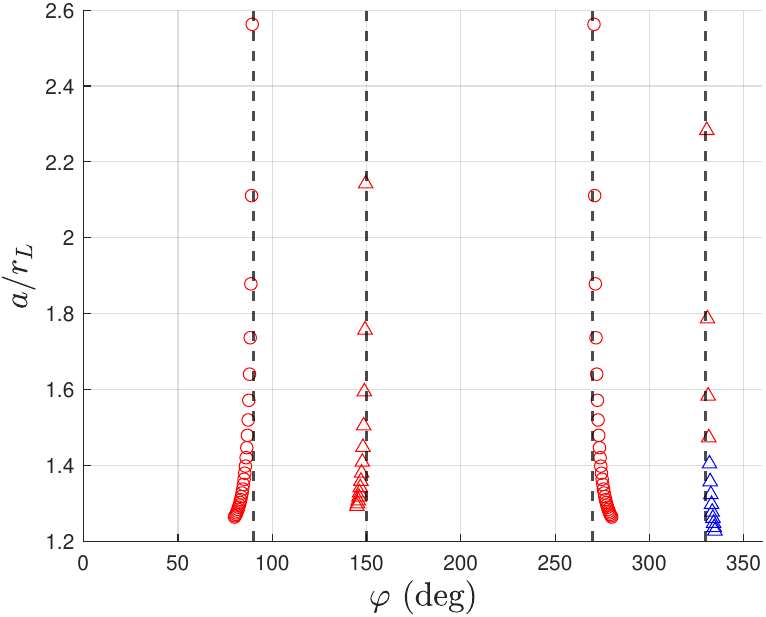}
\includegraphics[width=0.45\textwidth]{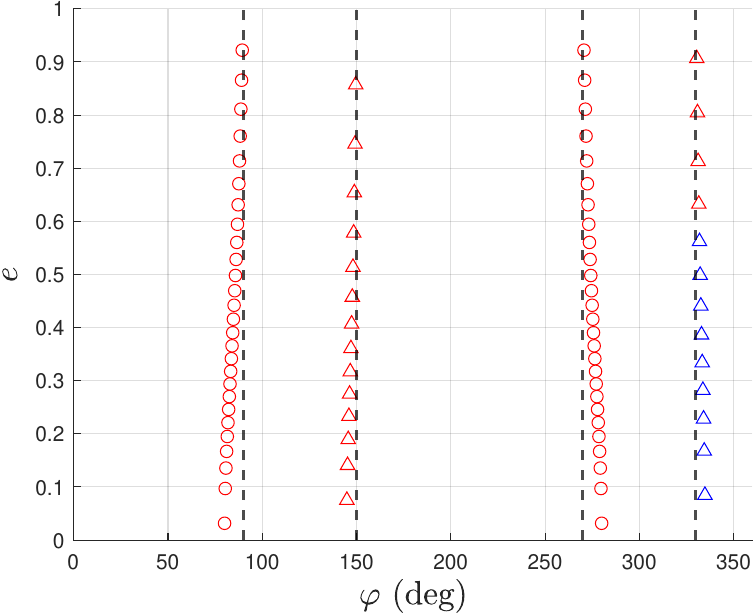}
\caption{Equilibrium states for $\theta=0^\circ$ (circles) and $\theta=60^\circ$ (triangles). Red markers denote unstable equilibria, while blue markers denote stable ones. The dashed black lines correspond to the limiting configuration where the orbital distance approaches infinity and the spin--orbit angle $\varphi$ approaches $90^\circ$. Left: the $a$--$\varphi$ relation. Right: the $e$--$\varphi$ relation.}\label{fig:eqi1}
\end{figure*}

Figure~\ref{fig:eqi1} shows the equilibrium solutions for the Laplace surface when the companion's orbital inclination is fixed at $\theta=0^\circ$ (circle) and $\theta=60^\circ$ (triangle) with $r_M/r_L=1.17$. Blue points indicate stable orbits and red unstable. The left panel shows particle's semi-major axis $a$ versus particle's orbital direction $\varphi$, and the right panel particle's eccentricity $e$ versus $\varphi$. We can see the two branches of solution confined in a narrow range as required by the torque balance. Moreover, as shown in the right panel, when particle's orbital plane approaches to be perpendicular to black-hole spin ($\varphi\to 90^\circ$), the orbital eccentricity approaches to be very high ($e\to1$), indicating that a near-orthogonal orientation corresponds to a highly eccentric orbit. Comparison between $\theta=0^\circ$ (circle) and $\theta=60^\circ$ (square) indicates that companion's inclination can shift and deform the equiblacklibrium branches.

\begin{figure*}
\centering
\includegraphics[width=0.45\textwidth]{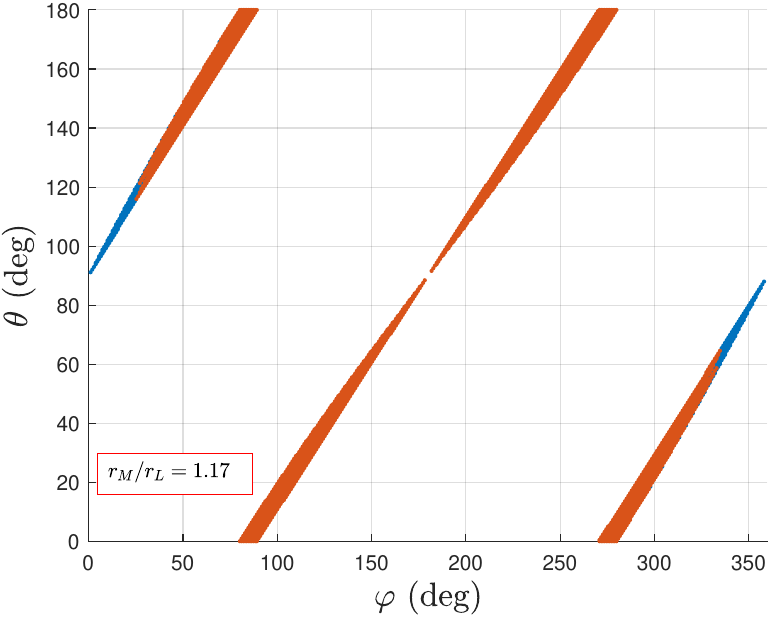}
\includegraphics[width=0.45\textwidth]{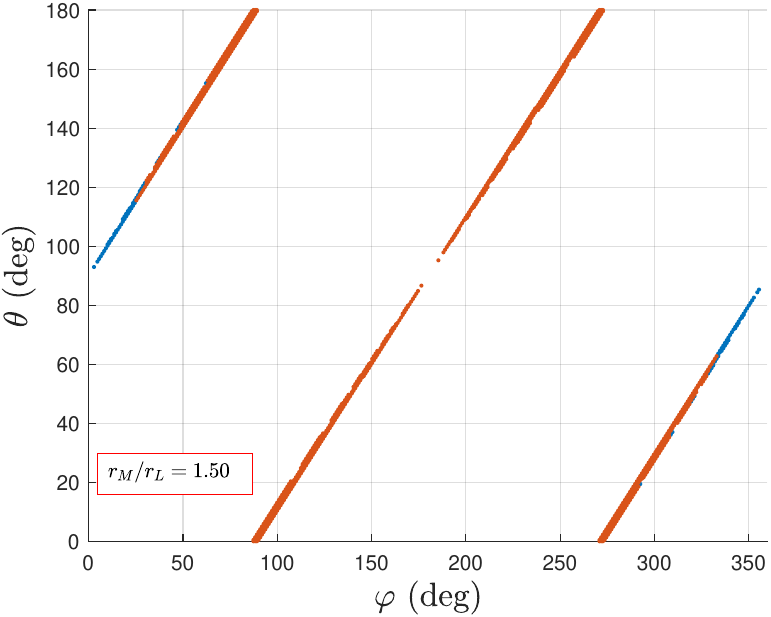}
\caption{Orbital stability in $\theta$--$\varphi$ space. Red denotes unstable equilibria and blue denotes stable ones. Left: $r_M/r_L=1.17$. Right: $r_M/r_L=1.5$.}\label{fig:stability}
\end{figure*}

We now evaluate the linear stability of the equilibrium solutions. The eigenvalue calculation method for linear stability analysis can be found in \S\ref{sec:equations}. Figure \ref{fig:stability} summarizes the stability of equilibrium states in the \((\theta,\varphi)\) plane, with the same color convention as in Figure \ref{fig:eqi1}. Equilibria exist only within a narrow strip where test-particle's orbits are nearly perpendicular to the companion's orbital plane, and this strip becomes progressively narrower as \(\theta\) approaches \(90^\circ\), i.e., companion's angular momentum becomes orthogonal to the black-hole spin. Stability is on the branch where the test-particle orbital direction tends to be parallel to the black-hole spin. Comparison between the two panels shows that the stable regime also depends on the ratio \(r_M/r_L\), i.e., a larger ratio produces a narrower band.

Moreover, the solutions display a reflection symmetry, i.e., if $(\theta,\varphi)$ is an equilibrium then so is its reflection to $(\theta=90^\circ, \varphi=180^\circ)$. To understand this, we examine Eqs.~(\ref{eq:j-precession}) and (\ref{eq:e-precession}) with two operators:  $R\circ\boldsymbol{j}$ denotes the reflection of $\boldsymbol{j}$ about the $x$–$z$ plane and $P\circ\boldsymbol{j}$ the reflection about the origin. Under the combined operations $R\circ\boldsymbol{j}$ and $RP\circ\boldsymbol{n_*}$, the right-hand side of the equations changes sign. Since an equilibrium state is the solution to the equation that the right-hand side is zero, changing its sign leaves the equilibrium unchanged. Therefore, the equilibrium states in the $(\theta,\varphi)$ plane are symmetric with respect to the point $(\theta=90^\circ,\varphi=180^\circ)$. Applying the same argument to the linearized perturbation equations Eqs.~(\ref{eq:perturbation_j}) and (\ref{eq:perturbation_e}), we find that the right-hand side remains invariant, and thus the stability exhibits the same symmetry.

\section{Black-hole effects on Lidov--Kozai cycles}
\label{sec:lk}

We now examine how relativistic precession modifies Lidov--Kozai cycles
driven by the companion's quadrupole potential. It is well known that
GR precession can suppress the eccentricity excitation in Lidov--Kozai cycles
\citep[e.g.][]{fabrycky2007shrinking,liu2015suppression}.
 In the parameter regime
\(a=0.004\)--\(0.02\,{\rm pc}\) and \(\chi=0.1\), the strength of Lense--Thirring term is small compared to
Schwarzschild apsidal precession term, thus we neglect Lense--Thirring term in our analytical calculation,
whereas retaining all terms in Eqs.~\eqref{eq:j-precession}--\eqref{eq:e-precession}
for our numerical integrations. The details of derivation are given in Appendix~\ref{appendix:zlk}, and below we give a brief summary about derivation.

When the Lense--Thirring term is neglected ($\bm J=0$), we can arbitrarily choose the coordinate system but usually companion's orbit as x-y plane. The test-particle angular momentum projected onto z axis $C\equiv\boldsymbol j\cdot\hat{\boldsymbol n}_*=j\cos I$ is conserved, where \(j=|\boldsymbol j|=\sqrt{1-e^2}\) and \(I\) is the
inclination to companion's orbit. We choose
initial $\bm e$ along the line of nodes for simplicity,
corresponding to \(\omega_0=0\), so that \(C=j_0\cos I_0\) with
\(j_0=\sqrt{1-e_0^2}\). Energy conservation then determines the maximum
eccentricity by
\begin{equation}
\label{eq:emax-equation}
\begin{aligned}
&-6\left(e_{\max}^2-e_0^2\right)
+15e_{\max}^2
\left(
1-\frac{C^2}{1-e_{\max}^2}
\right)
\\
&\hspace{2.0cm}
=
24\eta
\left[
\frac{1}{\sqrt{1-e_{\max}^2}}
-\frac{1}{j_0}
\right],
\end{aligned}
\end{equation}
where
$\eta
\equiv
GM/(ac^2)/\epsilon_*
\propto a^{-4}$.
In the classical limit \(\eta=0\) and \(e_0=0\),
Eq.~\eqref{eq:emax-equation} reduces to the standard Lidov-Kozai result
$e_{\max}^2
=
1-(5/3)\cos^2 I_0$.
The relativistic term therefore suppresses the eccentricity excitation,
with stronger suppression at smaller $a$ (note that $\eta\propto a^{-4}$). The same energy equation also determines the period of the eccentricity magnitude \(e=|\boldsymbol e|\). The eccentricity-oscillation period can be derived as
\begin{equation}
\label{eq:lk-period}
\begin{aligned}
T_e
={}&
\frac{8}{n\epsilon_*}
\int_{j_{\min}}^{j_0}
\frac{dj}{A(j)}\left\{
S(j)[1-S(j)]
\right\}^{-1/2}.
\end{aligned}
\end{equation}
Here \(j_{\min}=\sqrt{1-e_{\max}^2}\). The  functions
\(A(j)\) and \(S(j)\) are defined in
Eqs.~\eqref{eq:app-A-B} and \eqref{eq:app-S} of
Appendix~\ref{appendix:zlk}. 

Figure~\ref{fig:lk-time} shows the numerical integrations of
Eqs.~\eqref{eq:j-precession}--\eqref{eq:e-precession}. At small \(a\),
Schwarzschild apsidal precession strongly suppresses the eccentricity
excitation. As \(a\) increases, the relativistic precession becomes weaker
and the companion induces progressively larger-amplitude Lidov--Kozai
oscillations. Figure~\ref{fig:lk-effect} compares the analytical predictions with the
numerical solutions of the full vector equations. The numerical period is
measured between successive maxima of \(e(t)=|\boldsymbol e(t)|\).
The good agreement between the analytical and numerical results shows
that the Lense--Thirring correction is indeed negligible. The eccentricity oscillation frequency varies non-monotonically with testparticle's semi-major axis $a$, whereas the oscillation amplitude is strongly suppressed at
small \(a\). The period reaches its maximum near the transition between the
relativistically suppressed regime and the companion-dominated regime.
Across this transitional region, the maximum eccentricity increases rapidly.

\begin{figure}
\centering
\includegraphics[width=0.46\textwidth]{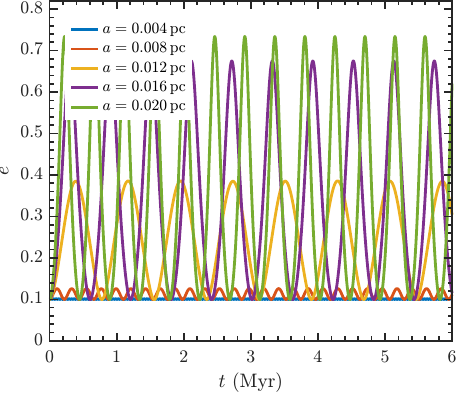}
\caption{Eccentricity magnitude \(e=|\boldsymbol e|\) obtained from direct integrations
of Eqs.~\eqref{eq:j-precession}--\eqref{eq:e-precession}.
The parameters are
\(M=4.15\times10^6M_\odot\),
\(M_*=10^5M_\odot\),
\(a_*=0.1\,{\rm pc}\),
\(e_*=0\),
\(e_0=0.1\),
\(I_0=60^\circ\),
and \(\chi=0.1\).
}
\label{fig:lk-time}
\end{figure}

\begin{figure*}
\centering
\includegraphics[width=0.45\textwidth]{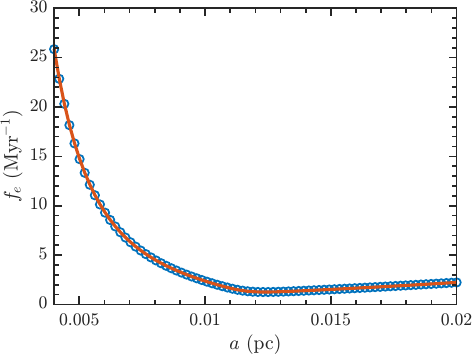}
\includegraphics[width=0.45\textwidth]{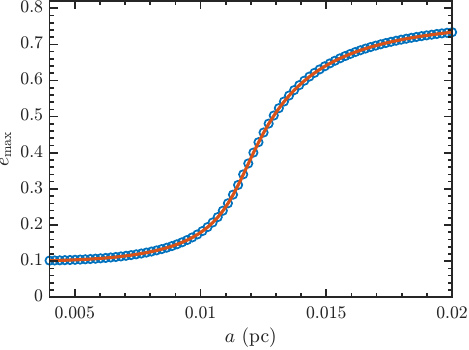}
\caption{Relativistic modification of the doubly averaged quadrupole
Lidov--Kozai cycles. Points are obtained from numerical integrations of
Eqs.~\eqref{eq:j-precession}--\eqref{eq:e-precession}
with \(\chi=0.1\). The curves are calculated from Eq.~\eqref{eq:lk-period} in the left panel
and Eq.~\eqref{eq:emax-equation} in the right panel.
Left: eccentricity-oscillation frequency
\(f_e=1/T_e\).
Right: maximum eccentricity reached during one complete cycle.
The remaining parameters are the same as in Fig.~\ref{fig:lk-time}.
}
\label{fig:lk-effect}
\end{figure*}

\section{Application to S-cluster}\label{sec:application}

Our model has two astrophysical consequences. Because the secular equilibria occupy a narrow range of orientations and most equilibrium branches are linearly unstable, a population of test particles or a gaseous ring around a Kerr black hole subject to a companion's potential will not precess in a well regulated manner but instead be in chaotic orbits. As an example, we apply to the S-cluster around the Galactic Center. There is a supermassive black hole in the center and a disk composed of stars which can be considered as a companion star in the outer region. Using a plausible spin range $\chi\in[0.1,0.9]$ and the orbital distribution \citep{genzel2010galactic} to constrain the other parameters, we obtain a lower limit on the ratio $r_M/r_L \gtrsim 2$. Consequently, the Laplace surface is narrow and typically unstable, i.e., orbits that initially belong to a partially coherent structure will be driven to a wide spread of inclination and ascending node. Observationally, the nearly isotropic orientation of the S-cluster could result from such secular evolution, while more distant stars remain confined in a coherent disk. The combined action of the stellar torque and Lense-Thirring precession may gradually erase an initially coherent stellar structure. On the other hand, \cite{fragione2020upper} argued that the S-stars are not isotropic but may instead trace two nearly perpendicular disks; if this configuration persists over the stellar lifetimes, the Lense-Thirring precession must be slow enough not to disrupt it, yielding an upper limit on the black-hole spin of $\chi\lesssim0.1$. Although this interpretation remains debated, it underscores the importance of spin constraint from the orbital architecture. In contrast, if a stellar cluster or ring close to a supermassive black hole exhibits little secular evolution, it may be residing on a stable Laplace surface, in which case the black-hole spin vector $\boldsymbol{J}$ could in principle be inferred from the orbital configuration.

\begin{figure*}
\centering
\includegraphics[width=0.45\textwidth]{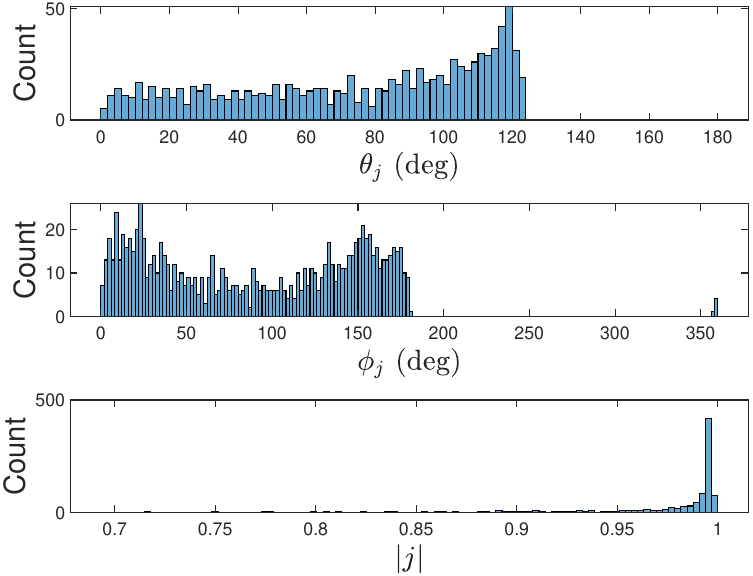}
\includegraphics[width=0.45\textwidth]{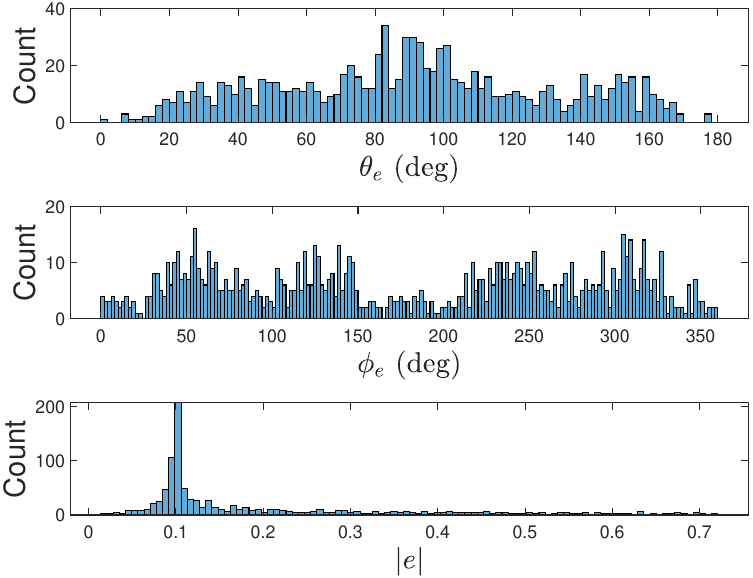}
\caption{$\theta_j$ and $\phi_j$ are the polar and azimuthal angles of the orbital angular momentum vector $\boldsymbol{j}$, while $\theta_e$ and $\phi_e$ are the polar and azimuthal angles of the eccentricity vector $\boldsymbol{e}$. The figure shows the distribution of $\boldsymbol{j}$ and $\boldsymbol{e}$ after 6~Myr of evolution with the black-hole spin $\chi=0.1$.}
\label{fig:application}
\end{figure*}

Figure \ref{fig:application} shows the results of numerical integration of the secular equations \eqref{eq:j-precession}--\eqref{eq:e-precession}, with the parameters \(\theta=60^\circ\), $M_{*}=10^4M_{\odot}$, $\tilde{a}_{\ast}=0.1\mathrm{pc}$ and $\chi=0.1$. Here we adopt the hypothesis that the S-stars formed in an inner disk whose angular momentum is aligned with the black-hole spin. Accordingly, the initial orbital angular-momentum vectors are taken to be parallel to the black-hole spin. The observed clockwise (outer) disk has an angular-momentum \(\boldsymbol{n_*}\) tilted by an angle \(\theta\) with respect to the black-hole spin. We therefore consider such an initial configuration, i.e., the S-star progenitors occupy an inner disk aligned with the black-hole spin while the observed clockwise disk forms a more extended structure misaligned with the black-hole spin. This configuration can arise naturally by the Bardeen–Petterson alignment. The initial S-star population is distributed in semimajor axis uniformly in the interval $a \in [0.004,\,0.02]\,\mathrm{pc}$. Each orbit is assigned an initial eccentricity $e_0=0.1$, and its initial angular-momentum vector is taken to be parallel to the black-hole spin axis. We integrate the secular equations over the typical S-star lifetime 6 Myr \citep{genzel2010galactic}. Our model can well interpret a nearly isotropic distribution of orbital orientation of S-cluster (top two rows), but cannot produce the high-eccentricity probably excited by dynamical scattering or density wave in the inner disk. We also tested some other values of spin $\xi$ and the results are similar. Therefore, the coupling between the black-hole spin and the stellar disk is shown to efficiently alter the orientation of S-cluster orbits. 

We need to notice that in the realistic situation the orbits are not test-particle trajectories, i.e., stellar gravitation can compete against the Lense-Thirring precession, and this effect is not considered in our model.

\section{Summary and discussions}\label{sec:summary}

In this paper, we have derived the secular, orbit-averaged equations governing a test particle orbiting a Kerr black hole under the tidal influence of a distant companion and used them to identify Laplace surface for eccentric orbits. Equilibrium states exist only in a narrow range of inclinations and appear on two distinct solution branches where particle's orbit is almost perpendicular to companion's orbit. A new characteristic radius $r_M$ naturally emerges alongside the classical Laplace radius $r_L$, and the ratio $r_M/r_L$ controls the branch width (with larger $r_M/r_L$ corresponding to a narrower branch). Linear stability analysis shows that most equilibrium states are unstable, while stable solutions are concentrated near stellar obliquity 90 degree where particle's angular momentum is approximately parallel to the black-hole spin. These results can be applied to well interpret the nearly isotropic orientation of S-cluster.

Laplace surface is usually used to study the gas disk structure. Our results show that a Kerr black-hole disk exhibits complicated dynamics. When particle's orbit approaches orthogonal to companion's orbit, high eccentricity will induce strong nonlinearity. The competition between Lense–Thirring and companion's tidal torque will yield substantial disk warps, such that gas orbit crossing and elliptical instability will generate shocks to enhance dissipation, tidal stripping, and even disk breaking into independently precessing rings, with attendant episodes of rapid accretion \citep{nixon2012tearing}. High eccentricity also raises sensitivity to secular chaos and strong apsidal precession, which can destabilize coherent structures and change angular-momentum transport. Therefore, hydrodynamic simulation will be the next study how a Kerr black-hole disk on Laplace surface leads to high eccentricity and inclination and yields warps and breaking.

\section*{Data Availability}
 
The data underlying this article are available in the article.

\section*{Acknowledgements}
We thank Prof. Dong Lai for his suggestion on studying the S-cluster and Prof. Zhoujian Cao for his advice on numerical integration.

\bibliographystyle{mnras}
\bibliography{paper}

\appendix

\section{Milankovitch equations and linear perturbation equations}\label{appendix}

In Hamiltonian mechanics, the time evolution of any physical quantity $f$ is given by Milankovitch equations
\begin{equation}
\frac{d f}{dt} = \{ f,H \}
\label{hamiltonian}
\end{equation}
where $\{\}$ denotes Poisson bracket. The orbit-averaged Hamiltonian $\langle H\rangle(\bm j,\bm e)$ depends only on $\boldsymbol{j}$ and $\boldsymbol{e}$ ($a$ is constant), so that the evolution equations become
\begin{equation}
\frac{d f}{dt} = \{ f,\boldsymbol{j} \} \cdot \nabla_{\boldsymbol{j}}\langle H\rangle + \{ f,\boldsymbol{e} \} \cdot \nabla_{\boldsymbol{e}}\langle H\rangle.
\end{equation}
Substituting $f$ with $\boldsymbol{j}$ or $\boldsymbol{e}$ yields Milankovitch equations
\begin{equation}\label{eq:Milankovitch}
\begin{aligned}
\frac{d \boldsymbol{j}}{dt} &= -\frac{1}{\sqrt{GM a}}\left(\boldsymbol{e}\times \nabla_{\boldsymbol{e}}\langle H\rangle + \boldsymbol{j}\times \nabla_{\boldsymbol{j}}\langle H\rangle\right),\\
\frac{d \boldsymbol{e}}{dt} &= -\frac{1}{\sqrt{GM a}}\left(\boldsymbol{j}\times \nabla_{\boldsymbol{e}}\langle H\rangle + \boldsymbol{e}\times \nabla_{\boldsymbol{j}}\langle H\rangle\right).
\end{aligned}
\end{equation}

The linear perturbation equations for stability analysis read
\begin{equation}\label{eq:perturbation_j}
\begin{aligned}
\frac{d\boldsymbol{j}_{1}}{dt} &=
\frac{3}{4}\epsilon_{\ast}(\boldsymbol{j}_{0} \cdot \hat{\boldsymbol{n}}_*)\, \boldsymbol{j}_{1} \times \hat{\boldsymbol{n}}_*
+ \frac{3}{4}\epsilon_{\ast}(\boldsymbol{j}_{1} \cdot \hat{\boldsymbol{n}}_*)\, \boldsymbol{j}_{0} \times \hat{\boldsymbol{n}}_* \\
& - \frac{15}{4}\epsilon_{\ast}(\boldsymbol{e}_{0} \cdot \hat{\boldsymbol{n}}_*)\, \boldsymbol{e}_{1} \times \hat{\boldsymbol{n}}_*
- \frac{15}{4}\epsilon_{\ast}(\boldsymbol{e}_{1} \cdot \hat{\boldsymbol{n}}_*)\, \boldsymbol{e}_{0} \times \hat{\boldsymbol{n}}_* \\
&+ 2\epsilon_{J}\, \hat{\boldsymbol{J}} \times \boldsymbol{j}_{1},
\end{aligned}
\end{equation}

\begin{equation}\label{eq:perturbation_e}
\begin{aligned}
\frac{d\boldsymbol{e}_{1}}{dt} &=
\frac{3}{4}\epsilon_{\ast}(\boldsymbol{j}_{0} \cdot \hat{\boldsymbol{n}}_*)\, \boldsymbol{e}_{1} \times \hat{\boldsymbol{n}}_*
+ \frac{3}{4}\epsilon_{\ast}(\boldsymbol{j}_{1} \cdot \hat{\boldsymbol{n}}_*)\, \boldsymbol{e}_{0} \times \hat{\boldsymbol{n}}_* \\
& - \frac{15}{4}\epsilon_{\ast}(\boldsymbol{e}_{0} \cdot \hat{\boldsymbol{n}}_*)\, \boldsymbol{j}_{1} \times \hat{\boldsymbol{n}}_*
- \frac{15}{4}\epsilon_{\ast}(\boldsymbol{e}_{1} \cdot \hat{\boldsymbol{n}}_*)\, \boldsymbol{j}_{0} \times \hat{\boldsymbol{n}}_* \\
&+ 3\epsilon_{M}\, \boldsymbol{j}_{0} \times \boldsymbol{e}_{1}
+ 3\epsilon_{M}\, \boldsymbol{j}_{1} \times \boldsymbol{e}_{0}
+ \frac{3}{2}\epsilon_{\ast}\, \boldsymbol{j}_{0} \times \boldsymbol{e}_{1}\\
&+ \frac{3}{2}\epsilon_{\ast}\, \boldsymbol{j}_{1} \times \boldsymbol{e}_{0} + 2\epsilon_{J}\, \hat{\boldsymbol{J}} \times \boldsymbol{e}_{1}  - \frac{6\epsilon_J}{1-e^2}(\boldsymbol{j}_{0}\cdot\hat{\boldsymbol{J}})\,\boldsymbol{j}_{1}\times \boldsymbol{e}_{0}\\
&- \frac{6\epsilon_J}{1-e^2}(\boldsymbol{j}_{0}\cdot\hat{\boldsymbol{J}})\,\boldsymbol{j}_{0}\times \boldsymbol{e}_{1}- \frac{6\epsilon_J}{1-e^2}(\boldsymbol{j}_{1}\cdot\hat{\boldsymbol{J}})\,\boldsymbol{j}_{0}\times \boldsymbol{e}_{0}.
\end{aligned}
\end{equation}

\section{Relativistic Lidov--Kozai cycle amplitude and period}
\label{appendix:zlk}

Here we derive the maximum eccentricity and oscillation period used in
Section~\ref{sec:lk}. We set \(\epsilon_J=0\) to neglect Lense-Tirring precession, while retaining the
Schwarzschild apsidal precession. The quadrupole Hamiltonian of
test-particle due to companion reads
\begin{equation}
\label{eq:app-stellar-H}
\Phi_*
=\Phi_0\big[15(\boldsymbol e\cdot\hat{\boldsymbol n}_*)^2
-6e^2+1-3(\boldsymbol j\cdot\hat{\boldsymbol n}_*)^2
\big],
\end{equation}
where
\begin{equation}
\Phi_0=
\frac{GM_*a^2}
{8a_*^3(1-e_*^2)^{3/2}}.
\end{equation}
In the absence of the spin term, the Hamiltonian is axisymmetric about
\(\hat{\boldsymbol n}_*\), and hence
\begin{equation}
C\equiv
\boldsymbol j\cdot\hat{\boldsymbol n}_*
=j\cos I
\end{equation}
is conserved, where $j\equiv|\boldsymbol j|=\sqrt{1-e^2}$. The argument of periapsis \(\omega\) measured relative
to companion's orbital plane satisifes
\begin{equation}
\label{eq:app-edotn}
(\boldsymbol e\cdot\hat{\boldsymbol n}_*)^2
=
(1-j^2)
\left(1-\frac{C^2}{j^2}\right)
\sin^2\omega.
\end{equation}
Substitution into Eq.~\eqref{eq:app-stellar-H} gives
\begin{equation}
\Phi_*(j,\omega)
=
\Phi_0
\left[
B(j)+A(j)\sin^2\omega
\right],
\end{equation}
where
\begin{equation}
\label{eq:app-A-B}
\begin{aligned}
A(j)&=
15(1-j^2)
\left(1-\frac{C^2}{j^2}\right),
\\
B(j)&=
1-6(1-j^2)-3C^2.
\end{aligned}
\end{equation}

The Schwarzschild contribution is obtained by requiring the Hamiltonian to
reproduce the apsidal precession rate
\begin{equation}
\dot{\omega}_{\rm GR}
=
\frac{3nGM}{ac^2j^2},
\end{equation}
where mean motion $n=\sqrt{GM/a^3}$. Since the generalised momentum conjugate to \(\omega\) is angular momentum \(\Lambda j\), where \(\Lambda=\sqrt{GMa}\) is the angular momentum of a circular orbit, the required orbit-averaged term is obtained by canonical equation $\partial\Phi_{\rm GR}/\partial(\Lambda j)=\dot\omega_{\rm GR}$, i.e.,
\begin{equation}
\Phi_{\rm GR}(j)
=
-\frac{3(GM)^2}{a^2c^2j}.
\end{equation}
The reduced Hamiltonian therefore becomes
\begin{equation}
\label{eq:app-total-H}
\Phi(j,\omega)
=
\Phi_0
\left[
B(j)+A(j)\sin^2\omega
-\frac{24\eta}{j}
\right],
\end{equation}
with
\begin{equation}
\eta\equiv
\frac{GM/(ac^2)}{\epsilon_*}.
\end{equation}

For fixed \(a\) and \(C\), the canonical equations read
\begin{equation}
\label{eq:app-canonical}
\begin{aligned}
\frac{dj}{dt}
&=
-\frac{n\epsilon_*}{8}
A(j)\sin2\omega,
\\
\frac{d\omega}{dt}
&=
\frac{n\epsilon_*}{8}
\left[
12j
+30\left(\frac{C^2}{j^3}-j\right)\sin^2\omega
+\frac{24\eta}{j^2}
\right].
\end{aligned}
\end{equation}
For the initial condition \(\omega_0=0\), we define
\begin{equation}
j_0=\sqrt{1-e_0^2},
\qquad
\mathcal E_0=
B(j_0)-\frac{24\eta}{j_0}.
\end{equation}
The high-eccentricity turning point satisfies \(dj/dt=0\) at
\(\omega=\pi/2\). Writing
\(j_{\min}=\sqrt{1-e_{\max}^2}\), conservation of
Eq.~\eqref{eq:app-total-H} gives
\begin{equation}
\label{eq:app-turning-energy}
B(j_{\min})-B(j_0)+A(j_{\min})=24\eta\left(\frac{1}{j_{\min}}-\frac{1}{j_0}\right).
\end{equation}
Using \(j_{\min}^2=1-e_{\max}^2\), this relation directly yields
Eq.~\eqref{eq:emax-equation} in the main text.

The period follows from the same energy curve. Equation
\eqref{eq:app-total-H} gives
\begin{equation}
\label{eq:app-S}
\sin^2\omega
\equiv S(j)
=
\frac{
\mathcal E_0-B(j)+24\eta/j
}{
A(j)
}.
\end{equation}
Differentiating the reduced Hamiltonian with respect to \(\omega\), by means of
\(\Phi_0/\sqrt{GMa}=n\epsilon_*/8\), gives
\begin{equation}
\frac{dj}{dt}
=
-\frac{n\epsilon_*}{8}
A(j)\sin2\omega.
\end{equation}
Since
\begin{equation}
|\sin2\omega|
=
2\sqrt{S(j)[1-S(j)]},
\end{equation}
the time required to move between the two turning points is
\begin{equation}\label{eq:app-T_e}
T_e
=\int dt
=\frac{8}{n\epsilon_*}
\int_{j_{\min}}^{j_0}
\frac{dj}{A(j)}
\left\{
S(j)[1-S(j)]
\right\}^{-1/2}.
\end{equation}
This is Eq.~\eqref{eq:lk-period} in the main text.

\bsp
\label{lastpage}

\end{document}